\documentclass[conference]{IEEEtran}

\usepackage[english]{babel}

\usepackage{color}
\usepackage[usenames,dvipsnames]{xcolor}

\usepackage{graphicx}
 \graphicspath{ {img/} } 
\usepackage[utf8]{inputenc}

\usepackage{setspace}
\usepackage{pgfplots}
\usepackage{array}
\usepackage{booktabs}
\pgfplotsset{compat=newest}
\usepackage{subcaption}

\usepackage{amssymb}
\usepackage{pifont}

\usepackage{xifthen}

\usepackage[inline,shortlabels]{enumitem} 
\newlist{inlinelist}{enumerate*}{1}
\setlist*[inlinelist,1]{%
  label=(\roman*),
}

\usepackage{etoolbox}

\usepackage{rotating}
\usepackage{adjustbox}

\usepackage{multicol}
\usepackage{multirow}
\usepackage{xspace} 

\usepackage{siunitx}
\usepackage{pgfplots}
\usepackage{tikz}
\usetikzlibrary{pgfplots.dateplot}
\usetikzlibrary{matrix,chains,positioning,decorations.pathreplacing}
\usetikzlibrary{patterns,calc,fit,arrows}
\usetikzlibrary{shapes.geometric}

\usepackage{pgfplotstable}
\usepackage{listings}
\definecolor{listingBG}{HTML}{FFFFCB}%
\definecolor{listingFrame}{HTML}{BBBB98}%
\definecolor{listingLineno}{rgb}{0.5,0.5,1.0}%

\usepackage{url}

\definecolor{LightGrey}{rgb}{0.975,0.975,0.975}
\definecolor{keyword}{HTML}{7F0055}

\lstdefinelanguage{java}{
        commentstyle=\color{Gray},
	morecomment=[l]{//},
	morecomment=[s]{/*}{*/},
	morestring=[b]",
        classoffset=0,
	morekeywords={public,private,static,final,class,extends,switch,case,break,finally,try,catch,void,int,boolean,throws,throw,return,if,else,new},
	keywordstyle=\color{keyword}\bfseries
}

\usepackage{hyperref}
\usepackage{cleveref}

\hypersetup{
  breaklinks   = true,
  colorlinks   = true, 
  urlcolor     = blue, 
  linkcolor    = blue, 
  citecolor    = red   
}
\hypersetup{final} 

\usepackage[final,nomargin,inline,index]{fixme} 
\fxusetheme{color}

\FXRegisterAuthor{bart}{anbart}{\color{magenta} {\underline{bart}}}
\FXRegisterAuthor{sergio}{ansergio}{\color{red} {\underline{sergio}}}

\newcommand{\ifempty}[3]{%
  \ifthenelse{\isempty{#1}}{#2}{#3}%
}

\newcommand{\hidden}[1]{}

\newcommand{\Real}[1]{\mathrm{Real}}

\newcommand{\codefont}{\fontsize{9}{9}\selectfont}
\newcommand{\code}[1]{{\tt\codefont {#1}}}

\def\etc{etc.\@\xspace}

\newcommand{\eg}{e.g.\@\xspace}
\newcommand{\ie}{i.e.\@\xspace}

  {}

\newcommand{\sig}[3][]{\mathit{sig}^{#1}_{#2}\ifempty{#3}{}{({#3})}}

\def\txColor{\color{MidnightBlue}}

\newcommand{\BTC}{\textup{%
  \leavevmode
  \vtop{\offinterlineskip 
    \setbox0=\hbox{B}%
    \setbox2=\hbox to\wd0{\hfil\hskip-.03em
    \vrule height .3ex width .15ex\hskip .08em
    \vrule height .3ex width .15ex\hfil}
    \vbox{\copy2\box0}\box2}}}

\newcommand{\USD}{\mbox{\textit{USD}}\xspace}
\newcommand{\USDfmt}[1]{\SI[round-precision=0,round-mode=places]{#1}}
\newcommand{\BTCfmt}[1]{\SI[round-precision=0,round-mode=places]{#1}}

\def\txColor{\color{MidnightBlue}}
\newcommand{\txFmt}[1]{{\txColor{\sf #1}}}

\newcommand{\tx}[2][]{\txFmt{#2}_{\txColor{#1}}}
\newcommand{\txT}[1][]{\tx[#1]{T}}

\newcommand{\txIn}[2][]{{\sf in}\ifempty{#1}{\ifempty{#2}{}{: {#2}}}{${}_{#1}$\ifempty{#2}{}{: {#2}}}}
\newcommand{\txWit}[2][]{{\sf wit}\ifempty{#1}{\ifempty{#2}{}{: {#2}}}{${}_{#1}$\ifempty{#2}{}{: {#2}}}}
\newcommand{\txOut}[3][]{{\sf out}\ifempty{#1}{\ifempty{#3}{}{({#2}): {#3}}}{${}_{#1}$\ifempty{#3}{}{({#2}): {#3}}}}
\newcommand{\txVal}[2][]{{\sf val}\ifempty{#1}{\ifempty{#2}{}{: {#2}}}{${}_{#1}$: {#2}}}

\DeclareMathAlphabet{\mathbfsf}{\encodingdefault}{\sfdefault}{bx}{n}

\def\pmvColor{\color{ForestGreen}}
\newcommand{\pmvFmt}[1]{{\pmvColor{\sf #1}}}
\newcommand{\pmv}[2][]{\pmvFmt{#2}_{\pmvColor{#1}}\xspace}
\newcommand{\pmvA}[1][]{\pmv[{#1}]{A}}
\newcommand{\pmvB}[1][]{\pmv[{#1}]{B}}
\newcommand{\pmvC}[1][]{\pmv[{#1}]{C}}

\crefname{appendix}{appendix}{appendices}
\Crefname{appendix}{Appendix}{Appendices}
\crefname{notation}{notation}{notations}
\Crefname{notation}{Notation}{Notations}

\definecolor{LightGrey}{rgb}{0.95,0.95,0.95}
\definecolor{keyword}{HTML}{7F0055}

\newlength\replength
\newcommand\repfrac{.1}

\newcommand\rulewidth{.6pt}
\newcommand\tdashfill[1][\repfrac]{\cleaders\hbox to \replength{%
  \smash{\rule[\arraystretch\ht\strutbox]{\repfrac\replength}{\rulewidth}}}\hfill}

\newcommand\tdotfill[1][\repfrac]{\cleaders\hbox to \replength{%
  \smash{\raisebox{\arraystretch\dimexpr\ht\strutbox-.1ex\relax}{.}}}\hfill}

\newcommand{\var}[2][]{#2_{#1}} 
\newcommand{\varX}[1][]{\var[#1]{x}} 
\newcommand{\varY}[1][]{\var[#1]{y}} 

\newcommand{\val}[2][]{#2_{#1}} 
\newcommand{\valV}[1][]{\val[#1]{v}}

\newcommand{\versig}[2]{{\sf versig}_{#1}({#2})}

\newcommand{\xmark}{\text{\ding{55}}}%

\newcommand{\Ponzi}{\textbf{P}\xspace}
\newcommand{\nonPonzi}{\textbf{nP}\xspace}
\newcounter{valTotPonzi}
\newcommand{\totPonzi}{\arabic{valTotPonzi}\xspace}

\newcounter{valTotCluster}
\newcommand{\totCluster}{\arabic{valTotCluster}\xspace}

\newcounter{valTotTxIn}
\newcounter{valTotTxTotal}
\newcommand{\totTxTotal}{\BTCfmt{\arabic{valTotTxTotal}}\xspace}

\newcounter{valTotInBTC}
\newcommand{\totInBTC}{\BTCfmt{\arabic{valTotInBTC}}\xspace}

\newcounter{valTotInUSD}
\newcommand{\totInUSD}{\USDfmt{\arabic{valTotInUSD}}\xspace}
\newcommand{\totInUSDapprox}{$\sim$\USDfmt{10}~millions\xspace}

\newcounter{numberFeatures}
\newcommand{\totFeatures}{\arabic{numberFeatures}\xspace}

\newcounter{numberTotalAddressesRandom}
\newcommand{\totRandomAddresses}{\arabic{numberTotalAddressesRandom}\xspace}

\newcounter{numberTotalAddresses}
\newcommand{\totAddresses}{\arabic{numberTotalAddresses}\xspace}

\newcommand{\urldataset}{\href{https://goo.gl/ToCho7}{\code{goo.gl/ToCho7}}\xspace}
\newcommand{\urlgithub}{\href{https://github.com/bitcoinponzi}{\code{github.com/bitcoinponzi}}\xspace}

\newtoggle{anonymous}
\togglefalse{anonymous}

\makeatletter
\renewcommand\paragraph{\@startsection{paragraph}{4}{\z@}%
  {2.25ex \@plus 1ex \@minus .2ex}%
  {-0.75em}%
  {\normalfont\normalsize\bfseries}}
\makeatother

\begin{document}

\title{Data mining for detecting Bitcoin Ponzi schemes}

\iftoggle{anonymous}{
\author{
  \IEEEauthorblockN{Given Name Surname}
  \IEEEauthorblockA{\textit{dept. name of organization (of Aff.)} \\
    \textit{name of organization (of Aff.)}\\                                       
    City, Country}
}
}{
\author{\IEEEauthorblockN{Massimo Bartoletti, Barbara Pes, Sergio Serusi}
	\IEEEauthorblockA{{\em University of Cagliari} \\ Cagliari, Italy \\
		Email: \texttt{\{bart,pes,serusisergio\}@unica.it}}
}
}

\maketitle

\begin{abstract}
  Soon after its introduction in 2009, Bitcoin has been adopted by cyber-criminals,
  which rely on its pseudonymity to implement virtually untraceable scams.
  One of the typical scams that operate on Bitcoin are the so-called Ponzi schemes.
  These are fraudulent investments which repay users with the funds invested by new users that join the scheme,
  and implode when it is no longer possible to find new investments.
  Despite being illegal in many countries, 
  Ponzi schemes are now proliferating on Bitcoin, 
  and they keep alluring new victims, who are plundered of millions of dollars.
  We apply data mining techniques to detect Bitcoin addresses related to Ponzi schemes. 
  Our starting point is a dataset of features of real-world Ponzi schemes,
  that we construct by analysing, on the Bitcoin blockchain, the transactions used to perform the scams.
  We use this dataset to experiment with various machine learning algorithms, 
  and we assess their effectiveness through standard validation protocols and performance metrics.
  The best of the classifiers we have experimented can identify most of the Ponzi schemes 
  in the dataset, with a low number of false positives.
\end{abstract}


\section{Introduction}
\label{sec:intro}

Bitcoin~\cite{bitcoin,Bonneau15ieeesp} is decentralized cryptocurrency,
which allows secure transfers of money --- the bitcoins --- 
without the intermediation of trusted authorities. 
All transfers of bitcoins are recorded on the \emph{blockchain}, 
an immutable public ledger of transactions 
maintained by a peer-to-peer network through a distributed consensus protocol.

Users can send and receive bitcoins without revealing their true identity:
rather, they use pseudonyms (called \emph{addresses}), 
which may even be generated fresh for each transaction.
Although several approaches to de-anonymise addresses have been proposed~\cite{Reid13analysis,Meiklejohn13imc,Meiklejohn16cacm,Moser17anonymous,Spagnuolo14bitiodine},
specular attempts to strengthen the anonymity of Bitcoin~\cite{Androulaki13fc,Bonneau14mixcoin,Moser17anonymous,Moser17cybersecurity,Ziegeldorf18fgcs}
reinforce the perception that criminal activities on Bitcoin are easy to implement, and hard to detect.

Besides classic criminal activities like ransomware~\cite{Spagnuolo14bitiodine,Liao16ecrime,Bistarelli18itasec}
and money laundering~\cite{Brenig15economic,Moser13inquiry}, 
Bitcoin is currently being used as a payment infrastructure for \emph{Ponzi schemes}~\cite{Vasek15fc}.
These are financial frauds disguised as ``high-yield'' investment programs:
actually, a Ponzi scheme repays users only with the funds invested by new users that join the scheme,
and so it implodes when it is no longer possible to find new investments~\cite{Artzrouni09,Moore12fc}.

Despite many victims are perfectly aware of their fraudulent nature,
and of the fact that they are illegal in many countries,
Bitcoin-based Ponzi schemes are proliferating. 
A recent study~\cite{Vasek18bw} inspects the posts on \url{bitcointalk.org}
(a popular discussion forum on Bitcoin), finding more than 
1800 Ponzi schemes from June 2011 to November 2016.
Estimating the economic impact of Bitcoin-based Ponzi schemes is more difficult,
due to the lack of datasets of Ponzi-related Bitcoin addresses:
a conservative estimate for the period from September 2013 to September 2014
shows that Ponzi schemes operated through Bitcoin have gathered more than $7$ millions \USD.
The absence of suitable prevention and intervention policies leads us to believe that
many other thousands of victims have been cheated since then, and plundered of millions of \USD.

Most of the existing approaches to the analysis of Bitcoin scams
require a laborious initial phase of manual or semi-automated search
on the web~\cite{Brenig15economic,Moser13inquiry,Vasek15fc,Vasek18bw,Moore13ijcip,Moser14fc,Slattery2014taking} 
in order to collect Bitcoin addresses involved in the scam.
Only after this phase it is possible to automatize the analysis,
\eg to quantify the impact of the scam by inspecting the associated 
transactions on the blockchain.
However, these approaches are ineffective when the scam addresses are not
publicly available, \eg because they are communicated privately to 
registered users, or published only through the deep web or the dark web.
In these cases it would be desirable to have tools that
\emph{automatically} search the Bitcoin blockchain for suspect behaviours, 
and identify the addresses associated to fraudulent activities.

Given the ever-increasing volumes of data to be managed
($\sim$300 millions of transactions, and several millions of distinct addresses) 
data mining techniques have become almost imperative for automatically extracting meaningful patterns for fraud detection. 
Outside the cryptocurrency realm,
several works in the literature have explored these techniques with data from credit card operations, 
either in a \textit{supervised} setting 
(which requires a set of labelled observations from the past)~\cite{BHATTACHARYYA2011,Whitrow2009} 
or through \textit{unsupervised} approaches 
(which look for anomalous data occurrences or \textit{outliers})~\cite{QUAH2008,Weston2008}. 
However, despite an increasing amount of research in the field, practical implementations are rarely reported, 
as recently pointed out in~\cite{CARNEIRO2017}. 
Furthermore, the scarcity of publicly available datasets leaves unanswered 
many questions about which is the best strategy to deal with specific real-world scenarios~\cite{DALPOZZOLO2014}. 

The extension of existing fraud detection methods to cyber-crime analysis in Bitcoin is an almost unexplored field. 
A few attempts have been recently made to detect anomalies in the Bitcoin transaction network by unsupervised learning 
approaches~\cite{PhamL16,MonamoMT16}, 
but no work exists, to the best of our knowledge, 
that investigates how to learn detection models for specific types of scams (such as Ponzi schemes).

  \paragraph*{Contributions}

We investigate data mining techniques to automatically detect and quantify Bitcoin Ponzi schemes,
following the supervised learning approach.

In the absence of publicly available datasets, 
our first step is to retrieve from the web a collection of Bitcoin addresses related to Ponzi schemes.
To this purpose we manually search the main discussion forums on Bitcoin (\eg, Reddit and \url{bitcointalk.org})
for advertisements of ``high-yield'' investment programs, that inevitably hide Ponzi schemes.
Then, we visit the websites through which Ponzi schemes are operated
(possibly recovering old snapshots through \href{https://archive.org/}{Internet Archive}), 
hunting for their Bitcoin addresses.
We expand our collection through a semi-automatic visit of the websites 
that are linked to Bitcoin addresses on \url{blockchain.info/tags}.
Following this methodology, we collect \totPonzi Bitcoin addresses which gather deposits from investors of Ponzi schemes. 

In many cases, Ponzi schemes use multiple addresses: 
actually, some of them provide the deposit address only upon registration, 
generating a fresh address for each new user.
In order to retrieve some of these addresses,
we apply a clustering procedure on the addresses in our collection,
using the ``multi-input'' heuristic~\cite{Reid13analysis}. 
By analyzing the obtained clusters, we find that 19 out of \totPonzi Ponzi schemes in our collection 
use more then one address, for a total of \totCluster addresses.
Overall, these clusters have received deposits for \totInUSDapprox \USD.

We then devise a set of features that can be useful to characterise Ponzi schemes.
These features range from simple statistics on the transactions to/from the clusters
(\eg, overall transferred value, ratio between incoming and outgoing transactions, \etc)
to more complex ones, 
like measures of inequality of the transferred values (\eg, Gini coefficients),
and measures of the activity of the scheme 
(\eg, lifetime, average delay between incoming and outgoing transactions, maximum number of daily transactions, \etc).
We extract from the Bitcoin blockchain the transactions 
of the clusters of addresses in our collection,
and we compute a dataset of features, which we make publicly available.
We complete this dataset with the features of \totRandomAddresses
randomly-chosen addresses.

We use this dataset to experiment with various supervised learning algorithms,
in order to automatically detect Ponzi schemes.
We formalise the detection model 
as a \textit{binary classification problem}, 
where the task is to distinguish between ‘Ponzi’ and ‘non-Ponzi’ 
class instances. 
One of the most critical challenges we had to face
is the \textit{class imbalance problem}, 
which is commonly encountered in fraud detection systems~\cite{ABDALLAH2016}. 
In a supervised learning setting, as the one here considered, this problem occurs when one class is very rare compared to the other(s), 
thus making hard to discover robust patterns for the minority class (like ``finding a needle in a haystack''). 
Indeed, classifiers are usually designed to minimize 
the total number of classification errors, 
and tend to be overwhelmed by the majority class.

In fraud detection applications, as in many domains with imbalanced class distributions, 
a correct classification of the rare class (\ie, the `Ponzi' class in our problem) 
is far more important than a correct classification of the majority class (\ie, the `non-Ponzi' class). 
The underlying assumption is that the cost of misclassifying a fraudulent 
case is much higher than the cost of misclassifying a legitimate case
(as the latter error can be corrected \emph{a posteriori} through a 
further analysis). 
In this work we experiment with the two main approaches proposed in the literature,
\ie \textit{sampling-based} approaches~\cite{Chawla04} and \textit{cost-sensitive} approaches~\cite{Thai-Nghe10}.

A number of experiments across different settings resulted in a detection model with good performance, 
which is finally applied, with promising results, to an independent set of data.
The supervised method Random Forest proved to be the most effective 
and most versatile one.
In our dataset, containing the features of \totAddresses clusters of addresses
(proportion of one fraud to 200 not fraud), 
Random Forest has obtained a Recall of 0.969 for Ponzi schemes, 
and it has classified correctly 31 Ponzi schemes out of \totPonzi.

\smallskip
In summary, our main contributions are:
\begin{enumerate}
\item a public dataset of addresses and features of Bitcoin Ponzi schemes
(\urldataset);
\item an open-source tool that extracts the dataset from the Bitcoin blockchain (\urlgithub);
\item a systematic evaluation and comparison of different learning strategies for classifying Bitcoin Ponzi schemes;
\item the evaluation of the best classifier (among those we have experimented with) 
  on an independent dataset, which manages to identify most of the Ponzi schemes 
  in the dataset, with a low number of false positives;
\item an estimate of which are the most discriminating features 
  for detecting Ponzi schemes on Bitcoin.
\end{enumerate}

\smallskip
The rest of this paper is organized as follows.
\Cref{sec:bitcoin} gives a minimalistic introduction to Bitcoin.
\Cref{sec:dataset} illustrates our methodology for collecting addresses of Ponzi schemes,
and for constructing a dataset of Ponzi-related features.
\Cref{sec:datamining} compares the effectiveness of various learning strategies.
Finally, \Cref{sec:conclusions} draws some conclusions.

\section{Bitcoin in a nutshell}
\label{sec:bitcoin}

In this~\namecref{sec:bitcoin} 
we give a short introduction to Bitcoin,
focussing on the notion that are needed later on for our
technical development. 

Bitcoin is a peer-to-peer infrastructure which allow users
to transfer currency --- the \emph{bitcoins} ($\BTC$).
Each Bitcoin user owns one or more pairs of asymmetric cryptographic keys:
public keys uniquely identify the user \emph{addresses},
while private keys are used to authorize payments.
Transfers of bitcoins are described by \emph{transactions}. 
The log of all transactions, 
is recorded on a public, immutable data structure (the \emph{blockchain}), 
determining the balance of each address. 
Users can receive bitcoins through different addresses: %
typically, addresses are generated fresh for each transaction,
to improve privacy.

The Bitcoin network is populated by a large set of nodes, called \emph{miners}, 
which collect transactions from users, 
and are in charge of appending the valid ones to the blockchain.
To this purpose, each miner keeps a local copy of the blockchain, 
and a set of transactions received by users.
Appending a new block of transactions to the blockchain
requires miners to solve a moderately-hard cryptographic puzzle,
which involves the transactions in the new block.
The difficulty of the puzzle is adjusted dynamically
to ensure that the average mining rate is of 1 block every 10 minutes.
The miner which solves the puzzle before the others
receives a reward in newly generated bitcoins (through the so-called \emph{coinbase} transactions), 
and a fee for each transaction included in the new block.

\begin{figure}[t]
  \resizebox{\columnwidth}{!}{
    \small
    \centering
    \begin{minipage}{1.1\columnwidth}
      \begin{tabular}{|l|}
        \hline
        \\[-8pt]
        \multicolumn{1}{|c|}{$\txT$} \\
        \hline
        \\[-6pt]
        \txIn[1]{$\cdots$} \\[1pt]
        \txWit[1]{$\cdots$} \\[1pt]
        \txOut[1]{$\varX$}{$\versig{\pmvA}{\varX}$} \\[1pt]
        \txVal[1]{$1 \BTC$} \\[1pt]
        \txOut[2]{$\varY$}{$\versig{\pmvB}{\varY}$} \\[1pt]
        \txVal[2]{$2 \BTC$} \\[1pt]
        \hline
      \end{tabular}
      \begin{tabular}{|l|}
        \hline
        \\[-8pt]
        \multicolumn{1}{|c|}{$\txT[\pmvA]$} \\
        \hline
        \\[-6pt]
        \txIn[1]{$(\txT,1)$} \\[1pt]
        \txWit[1]{$\sig{\pmvA}{\txT[\pmvA]}$} \\[1pt]
        \txOut[1]{$\varX$}{$\versig{\pmvB}{\varX}$} \\[1pt]
        \txVal[1]{$0.9 \BTC$} \\[1pt]
        \txOut[2]{$\varY$}{$\versig{\pmvA}{\varY}$} \\[1pt]
        \txVal[2]{$0.1 \BTC$} \\[1pt]
        \hline
      \end{tabular}
      \begin{tabular}{|l|}
        \hline
        \\[-8pt]
        \multicolumn{1}{|c|}{$\txT[\pmvB]$} \\
        \hline
        \\[-6pt]
        \txIn[1]{$(\txT,2)$} \\[1pt]
        \txWit[1]{$\sig{\pmvB}{\txT[\pmvB]}$} \\[1pt]
        \txIn[2]{$(\txT[\pmvA],1)$} \\[1pt]
        \txWit[2]{$\sig{\pmvB}{\txT[\pmvB]}$} \\[1pt]
        \txOut[1]{$\varX$}{$\versig{\pmvC}{\varX}$} \\[1pt]
        \txVal[1]{$2.5 \BTC$} \\[1pt]
        \hline
      \end{tabular}
    \end{minipage}
  } 
  \caption{Three Bitcoin transactions.}
  \label{fig:bitcoin}
\end{figure}

To explain how transactions work, we consider the example in~\Cref{fig:bitcoin},
which graphically represents three transactions. 
Each transaction has four (indexed) fields: $\txIn{}$, $\txWit{}$, $\txOut{}{}$, and $\txVal{}$. 
The $\txIn{}$ field  (for \emph{input}) contains a reference
to the transaction output to redeem.
The $\txWit{}$ field contains a piece of information called \emph{witness}, discussed below.
The field $\txOut{}{}$ contains an \emph{output script}:
intuitively, this is a predicate on one or more arguments, 
the actual values of which are provided by the witness of the redeeming transaction.
Finally, the field $\txVal{}$ determines the amount of bitcoins to be transferred.

Consider first the transaction $\txT$ in our example, 
where we neglect the fields $\txIn{}$ and $\txWit{}$ 
(\eg, we could assume that $\txT$ is a coinbase transaction).
The output at index $1$ allows to transfer $1 \BTC$ to user $\pmvA$:
namely, the output script $\versig{\pmvA}{\varX}$ 
verifies the signature $\varX$ of $\pmvA$ on the redeeming transaction.
Similarly, the output at index $2$ allows to transfer $2 \BTC$ to $\pmvB$.
Assume that the blockchain contains $\txT$, 
and that both its outputs are not redeemed by subsequent transactions.

Transaction $\txT[\pmvA]$ has one input, represented as the pair $(\txT,1)$,
meaning that it wants to redeem the output at index 1 of transaction $\txT$.
To do so, $\txT[\pmvA]$ carries in its witness a signature of $\pmvA$,
which is computed on the whole transaction $\txT[\pmvA]$
(\emph{except} for the $\txWit{}$ field itself).
This witness makes the output script $\versig{\pmvA}{\varX}$ in $\txT$ evaluate to true.
Therefore, when $\pmvA$ appends $\txT[\pmvA]$ to the blockchain,
it redeems the first output of $\txT$,
making available $0.9 \BTC$ for $\pmvB$, and keeping $0.1 \BTC$ for herself.

The transaction $\txT[\pmvB]$ has two inputs,
meaning that it wants to simultaneously redeem the second output of $\txT$
and the first output of $\txT[\pmvA]$.
Since these outputs are still unspent, and the witnesses in $\txT[\pmvB]$ 
satisfy the corresponding output scripts, 
then $\pmvB$ can append $\txT[\pmvB]$ to the blockchain,
making available $2.5 \BTC$ to user $\pmvC$.
The difference of $0.4 \BTC$ between the the sum of the values of all the inputs of $\txT[\pmvB]$ 
and the sum of the values of its outputs is paid as a fee to the miners.

Bitcoin transactions may be more general than the ones illustrated by the previous example.
For instance, the output script
is a program in a (not Turing-complete) scripting language, 
featuring a limited set of logic, arithmetic, and cryptographic operators.
Transactions can also specify time constraints on when they can be appended to the blockchain,
and also on when the redeeming transactions can be appended.
We omit a detailed presentation of these advanced features, 
since they are not required in the following sections.

\section{Dataset construction}
\label{sec:dataset}

The first step of our work is to collect Bitcoin addresses
through which Ponzi schemes receive money from investors 
(\Cref{sec:dataset:collection}).
We apply a clustering algorithm to the collected addresses,
finding that some schemes use wallets of hundreds of addresses
(\Cref{sec:dataset:clustering}).
We then devise a set of features that are relevant to the classification
of Ponzi schemes
(\Cref{sec:dataset:features}),
and we compute the values of these features on our clustered addresses,
obtaining a dataset that we use in~\Cref{sec:datamining} to train classifiers
(\Cref{sec:dataset:construction}).

  \subsection{Collection of Bitcoin addresses used by Ponzi schemes}
\label{sec:dataset:collection}

We perform a manual search on Reddit and \url{bitcointalk.org}, 
the main discussion forums on Bitcoin.
In particular, we focus on the subforum
\href{https://bitcointalk.org/index.php?board=207.0}{Gambling: Investor-based games}
of bitcointalk.org, 
where fraudsters are used to advertise Ponzi schemes as 
``high-yield investment programs'' (HYIP), or as gambling games.
Only in a few cases these advertisements explicitly include 
the Bitcoin address where to deposit money;
in all the other cases, to obtain the address we have to visit the websites 
where the Ponzi schemes are hosted.
However, many of these websites are no longer online: 
in such case we try to recover their snapshots through 
\href{https://archive.org/}{Internet Archive}.
For each website (either live or snapshot), we manually search 
its pages to find the deposit addresses 
(typically, different ``investment plans'' use different addresses).
Some websites only allow registered users to read the deposit address:
in these cases, we create an account, providing fake data.

We extend our search by considering all the addresses listed
on \href{https://blockchain.info/tags}{blockchain.info/tags},
a website which allows users to tag Bitcoin addresses.
Most of the tagged addresses also contain a link to the website 
where they are mentioned.
We develop a crawler to automatically parse these websites,
and rank them according to the number of Ponzi-related words contained in their pages.
To this purpose we use a dictionary, containing words like \eg
``Ponzi'', ``profit'', ``HYIP'', ``multiplier'', ``investment'', ``MLM''.
The crawler parses over 1500 websites (related to $\sim$3500 tags),
finding that $\sim$900 of them contain some Ponzi-related word. 
However, many of these sites ($\sim$600) are no longer accessible, even through~\href{https://web.archive.org/}{Internet Archive}. 
For the remaining websites, we manually search for deposit addresses,
creating fake accounts when needed. 

Overall, we find $\totPonzi$ deposit addresses of Ponzi schemes,
that we display in~\Cref{fig:dataset:collection}. 
Note that, while some Ponzi schemes use a single deposit address throughout their lives,
some others use multiple addresses, 
possibly generating a fresh address for each user (or set of users).
Address clustering, discussed below, allows us to
recover some of these addresses.

\begin{table}[t!]
  \centering				
  \small
  \caption{Collection of addresses of Bitcoin Ponzi schemes.}
  \resizebox{\columnwidth}{!}{
    \begin{tabular}{ll}  
      \toprule
      \textbf{Ponzi scheme} & \textbf{Deposit address} \\
      \midrule
      Nanoindustryinv.com  & 1Ee9ZiZkmygAXUiyeYKRSA3tLe4vNYEAgA\\
      GrandAgoFinance & 1MzNQ7HV8dQ6XQ52zBGYkCZkkWv2Pd3VG6\\	
      Cryptory &1FyedPPk923wRfmVphV1CLt3bVLGxHZXpK\\
      Leancy & 145SmDToAhtfcBQhNxfeM8hnS6CBeiRukY\\
      Minimalist10 &1FuypAdeC7mSmYBsQLbG9XV261bnfgWbgB\\	
      MiniPonziCoin &1F8ZKpjMDpnpF79mZ1pxZRoNKZgXm4Tf1d\\  
      120cycle &1E5MCTtXn7n2svpZ1bDHZXndY9K7qQeqZP\\ 
      10PERCENTBTC &1BtcBoSSnqe8mFJCUEyCNmo3EcF8Yzhpnc\\
      btcgains&	1PayoutRrC8wxxZ9ygmeaRj3qTPug8tDYu\\
      PonziIO   &1ponziUjuCVdB167ZmTWH48AURW1vE64q\\ 
      LaxoTrade &1LaxoTrQy51LnB289VmoSAgN6J6UrJbfL9\\
      OpenPonzi &	1BmZW65ZoeLa1kbL9MPFLfkS818mqFUSma\\
      BTC-doubler.us&	1AQp51H22WHDzLgK64NoUo3Bg3T183QR22\\
      BTC-doubler.com	&178BzARKjkszrTyx4TxBKHhzGLZijdE26e\\
      investorbitcoin.com&	1CpVAEg4BgVzjiHshgeZfitZLV1t1zo6Qg\\
      Ponzi120 & 12PoNZiEtabwkCU4YFffshWNF1cRiAk5nq\\
      RockwellPartners & 139eeGkMGR6F9EuJQ3qYoXebfkBbNAsLtV\\
      Twelverized &114Ap9G5nu78vESC648amPwSeqUorPtV5L\\
      CRYPTOSX2&	19YZYfMB3mfX8AixzV7aLqXuViDcntrfcK\\
      1hourbtc.pw	&1BsjsaHST2Qohs8ZHxNHeZ1UfWhtxoKHEN\\
      bitcoindoubler.fund&	1FNtgGsHhymmEUMXrMiFeMtZbuagnnS59c\\
      doublebitcoin.life&	1zmeu5BeWBprWyPv5ntNZKR7uThXaG9ic\\
      bitcoincopy.site123.me&	1EaSVdRuzcz4yjnTmibabyyrczvaQS8hAJ\\
      bitcoinprofit2&	1AXTqWYz1Bd3LZnq1Zf9vsgFBpqrKkHopx\\
      invest4profit	&1PZ8E5oT7EUVgEVz1Ggc7bjXe2byxr7wxG\\
      1getpaid.me&	1GetPaiDxjEuWN3KJTnY9Cbqv9QcR8zcME\\
      Ponzi.io (change)&	14ji9KmegNHhTchf4ftkt3J1ywmijGjd6M\\
      igjam.com&	1AQxcdPgMTTQghPXt1EXHU8vEjSn2kYrPQ\\
      7dayponzi&	195o79saDhUNHJ4DeMBYMekLmrQ848APxA\\
      world-btc.online&	1A88teD6QqXRHBMCyCkoxxBQHpJAztUz6e\\
      bestdoubler.eu&	13NZxtAnKk5mbCUHpxHqKwWTDJzFHMGHLh\\
      bitcoindoubler.prv.pl &	18Smkvyf3gJN4z59FhjJsCu6NhSYmZkNvG\\
      \bottomrule
    \end{tabular}
  } 
  \label{fig:dataset:collection}
\end{table}

  \subsection{Address clustering}
\label{sec:dataset:clustering}

Many techniques to break the anonymity of Bitcoin users
have been proposed in the literature.
This is achieved either
by grouping together (\emph{``clustering''}) the addresses controlled 
by each user~\cite{Reid13analysis,Meiklejohn13imc,Meiklejohn16cacm}, 
or by using observations on the underlying peer-to-peer 
network~\cite{Biryukov15sp,Dupont15codaspy},
or by combining both techniques~\cite{Neudecker17fc}.

Several heuristics for address clustering have been proposed
over the years.
Besides analysing the shape of the transaction graph,
some heuristics also take into consideration the behavior 
of standard clients~\cite{Meiklejohn16cacm}.
To construct our dataset of Ponzi-related addresses,
we use the \emph{multi-input} heuristic~\cite{Reid13analysis,Meiklejohn13imc},
the simplest and most efficient one.
The key assumption of this heuristic is that,
in a multi-input transaction (like \eg $\txT[\pmvB]$ in~\Cref{fig:bitcoin}), 
all the addresses referred to within the inputs
are controlled by the same user. 
These transactions occur, for instance, 
when a user $\pmvA$ wishes to transfer a certain amount of $\valV \BTC$
to another user $\pmvB$,
but none of the transactions in $\pmvA$'s wallet has an unspent output
of at least $\valV \BTC$. 
In this case, to avoid paying multiple transaction fees,
$\pmvA$ can perform the transfer in a single shot,
by putting on the blockchain a multi-input transaction 
redeemable by $\pmvB$,
where the sum of the values redeemed by the inputs is at least $\valV \BTC$.
Typical Bitcoin clients implement this by choosing the input transactions 
from $\pmvA$'s wallet, satisfying the assumption of the 
multi-input heuristic.

We show in~\Cref{fig:dataset:clustering} some statistics on the 
clusters that we obtain after applying the multi-input heuristic 
to the \totPonzi addresses in our collection.
The columns display the size of the clusters,
the overall number of transactions (either incoming or outgoing),
and the overall inflow, both in \BTC\xspace and in \USD.
To convert the amount of each transaction to \USD, 
we use the average exchange rate on the day of the transaction,
obtained from~\href{https://www.coindesk.com/price/}{www.coindesk.com/price}.
Overall, the Ponzi schemes in our collection gathered almost $10$ millions \USD;
the scheme that raised the most is Cryptory, 
with \mbox{$\sim 4.6$} millions \USD.

\begin{table}[t!]
	\centering				
	\small
	\caption{Top-10 Ponzi schemes by cluster size.}
	\resizebox{\columnwidth}{!}{
		\begin{tabular}{lrrrrrr}
			\toprule
			\textbf{Ponzi scheme} & \textbf{{\#}Addr.} & \textbf{{\#}Tx}
			& \textbf{In (\BTC)} & \textbf{In (\$)}  \\ 
			\midrule
			LaxoTrade           & 491 & \USDfmt{4798}  & 1,580 & \USDfmt{570106}{}   \\
			Cryptory            & 232 & \USDfmt{22823} & 9,439 & \USDfmt{4658008}{}  \\
			1hourbtc            & 180 & \USDfmt{1262}  & 36    & \USDfmt{42668}{}    \\
			120cycle            & 78  & \USDfmt{284}   & 14    & \USDfmt{8263}{}     \\
			bitcoindoubler.fund & 63  & \USDfmt{1143}  & 90    & \USDfmt{288849}{}   \\ 
			world-btc.online    & 41  & \USDfmt{302}   & 1     & \USDfmt{2060}{}     \\
			Ponzi.io            & 33  & \USDfmt{6311}  & 370   & \USDfmt{258368}{}   \\
			btcgains            & 14  & \USDfmt{789}   & 72    & \USDfmt{33246}{}    \\
			10PERCENTBTC        & 13  & \USDfmt{10077} & 107   & \USDfmt{42894}{}    \\
			investorbitcoin.com & 11  & \USDfmt{672}   & 312   & \USDfmt{158569}{}   \\
                  \midrule 
                  \textbf{Total (\totPonzi schemes)} & \totCluster & \totTxTotal & \totInBTC & \totInUSD \\
                  \bottomrule 
		\end{tabular}
	}
	\label{fig:dataset:clustering}
\end{table}
  \subsection{Features extraction}
\label{sec:dataset:features}

We now introduce a set of features, which are relevant for the classification of Bitcoin addresses.
\begin{itemize}

\item The \emph{lifetime} of the address, expressed in number of days. 
  This is computed as the difference between the date of the first transaction to the address,
  and the date of the last transaction to/from the address.
  
\item The \emph{activity days}, 
  \ie the number of days in which there has been at least a transaction to/from the address.
  
\item The maximum number of daily transactions to/from the address.

\item The \emph{Gini coefficient} of the values transferred to (resp.\ from) the address. 
Gini coefficients are a standard representation of the degree of inequality of wealth:
0 indicates perfect equality, while 100 is perfect inequality~\cite{Vasek15fc}.

\item The sum of all the values transferred to (resp.\ from) the address.

\item The number of incoming (resp.\ outgoing) transactions which transfer 
money to (resp.\ from) the address.

\item The ratio between incoming and outgoing transactions to/from the address.
  
\item The average (resp.\ standard deviation) of the values transferred to/from the address.
  
\item The number of different addresses which have transferred money to 
the address, and subsequently received money from it.
 
\item The minimum (resp.\ maximum, average) delay 
  between the time when the address has received some bitcoins, 
and the time it has sent some others.
  
\item The maximum difference between the balance of the address in two consecutive days.
  
\end{itemize}

All the features above are defined ``pointwise'' on single Bitcoin address.
We extend them to clusters in the straightforward way:
any feature on a cluster is the composition of the pointwise features on the addresses included in the cluster.
As an additional ``componentwise'' feature, we consider the number of addresses included in the cluster.

  \subsection{Dataset construction}
\label{sec:dataset:construction}

We construct a binary dataset that contains two classes of instances: 
Ponzi schemes (denoted as \Ponzi) and others (denoted as \nonPonzi). 
Each instance in the dataset corresponds to a cluster of Bitcoin addresses
(computed as shown in~\Cref{sec:dataset:clustering}),
and it is represented as a tuple of features 
(the ones defined in~\Cref{sec:dataset:features}, plus the class label \Ponzi or \nonPonzi). 
To compute these dataset instances we exploit an open-source tool for custom blockchain analytics~\cite{Bartoletti17serial}.

We populate the dataset with \totPonzi instances of the class \Ponzi 
(corresponding to our clusters of Ponzi schemes),
and with \totRandomAddresses randomly-chosen instances of the class \nonPonzi
(which are clustered with the multi-input heuristic as well).
This strong imbalance between the two classes 
(approximately, 1 Ponzi instance every 200 instances of non-Ponzi)
is needed to properly model the fact that Ponzi-related addresses
are extremely rare, compared to non-Ponzi ones.
Although 1/200 is still much higher then the expected ratio between Ponzi addresses and non-Ponzi ones,
it is a necessary compromise to meaningfully represent also the rare class in the dataset.

\section{Data mining for Ponzi schemes}
\label{sec:datamining}

We formalise the induction of a detection model for Bitcoin Ponzi schemes 
as a \textit{binary classification problem}, 
where the task is to distinguish between `Ponzi' and `non-Ponzi' 
class instances. 
The strategies for dealing with the imbalanced distribution 
of the classes are discussed below 
(\Cref{sec:datamining:imbalance}), 
along with the learning algorithms applied to induce the model 
(\Cref{sec:datamining:classifiers}), 
the performance metrics and the validation protocol 
(\Cref{subsec:validation}). 
A number of experiments across different settings 
(\Cref{sec:datamining-results}) 
resulted in a detection model with good performance, 
which is then applied, with promising results, 
to an independent set of data 
(\Cref{sec:datamining:application}).
Finally, we investigate which are the most relevant features, 
among those in our set, to detect Ponzi schemes 
(\Cref{sec:datamining:ranking}).
  \subsection{Class imbalance problem}
\label{sec:datamining:imbalance}

The \textit{class imbalance problem} is one of the most critical issues faced by fraud detection systems~\cite{ABDALLAH2016}. 
In a supervised learning setting, as the one here considered, this problem occurs when one class is very rare compared to the other(s), 
thus making hard to discover robust patterns for the minority class. 
Indeed, classifiers are usually designed to minimize the total number of classification errors, 
and tend to be overwhelmed by the majority class.

In fraud detection applications, as in many domains with imbalanced class distributions, 
a correct classification of the rare class (\ie, the \Ponzi class in our problem) 
is far more important than a correct classification of the majority class 
(\ie, the \nonPonzi class), 
since the cost of misclassifying a fraudulent case is usually
higher than the cost of misclassifying a legitimate case 
(as the latter can be corrected through \emph{ex-post} analyses). 

A number of approaches have been proposed in the literature for handling this problem, 
including \textit{sampling-based} approaches~\cite{Chawla04} and \textit{cost-sensitive} approaches~\cite{Thai-Nghe10}.

\paragraph*{Sampling-based approaches}
The basic idea is to modify the distribution of instances 
so that the minority class is adequately represented in the dataset used for model development. 
The most common sampling technique is \textit{random undersampling} (RUS), 
which consists in removing observations at random from the majority class. 
An alternative approach is \textit{random oversampling} (ROS), where some of the minority instances are replicated, 
but with an increased risk of overfitting, particularly with noisy data~\cite{DALPOZZOLO2014}. 
Though more sophisticated (and expensive) approaches exist, 
they have not proved to be superior in severe imbalance settings~\cite{Hulse07}. 
Furthermore, the effectiveness of sampling techniques may be dependent on the learning algorithm used 
(and on the adopted performance measure); as well, the extent of sampling for best performance may be domain-dependent.

\paragraph*{Cost-sensitive approaches}
Cost-sensitive learning involves the use of a \textit{cost matrix} 
which encodes the penalty of classifying instances from one class as another. 
In a class imbalance setting, where the focus is usually on rare instances, 
a misclassification of the minority class is penalized more than a misclassification of the majority class. 
For a given classification model, penalty terms are then used to derive an \textit{overall cost} 
which reflects the “weight” of the different types of classification errors, besides their total number. 
Cost-sensitive classification techniques take this cost matrix into consideration in the training phase, 
in order to generate the classification model with the lowest overall cost.

\medskip
In this work we consider both sampling-based and cost-sensitive approaches. 
Furthermore, we experiment with learners whose inner design can cope, at least to some extent, with imbalanced class distributions, 
as in the case of the RIPPER algorithm proposed in~\cite{Cohen95}. 

  \subsection{Classifiers}
\label{sec:datamining:classifiers}

In the induction stage of our detection model, 
we exploited \textit{RIPPER}, \textit{Bayes Network} and \textit{Random Forest} classifiers,
which are representatives of quite different learning strategies. 

\smallskip
\textbf{\textit{RIPPER}} is a propositional rule learner that relies on a \textit{sequential covering} logic~\cite{Cohen95}
to extract classification rules directly from training data. Rules are grown in a greedy fashion, 
starting from empty rule antecedents and repeatedly adding conjuncts in order to maximize the information gain measure. 
An incremental \textit{reduced error pruning} technique is used the refine the resulting rules. 
Since the algorithm is designed to give higher priority to the least frequent class, 
this approach is particularly suited for dealing with imbalanced classification tasks, 
as in the case of fraud detection.

\smallskip
\textbf{\textit{Bayes Network}} is a probabilistic model that represents, in the form of a \textit{directed acyclic graph}, 
the relation of conditional dependence among a set of variables 
(the features and the target class, in the context of classification problems). 
Probabilistic parameters are encoded in a set of tables, one for each node of the network, 
in the form of local conditional distributions of a variable given its parents. 
Both the network structure and the probability values can be estimated from a training set of labelled instances. 
Bayesian models have been applied in the context of fraud detection systems, \eg in~\cite{Maes02}.

\smallskip
\textbf{\textit{Random Forest}} is an ensemble method that exploits multiple decision trees built from random variants of the same data ~\cite{Breiman2001}. 
Although a single tree may be unstable and overly sensitive to the specific composition of the training set, 
the aggregation of the predictions made by the individual trees in the forest has been shown to achieve much better performance. 
Compared to other ensemble approaches, \textit{Random Forest} is computationally efficient 
and has proved to be a “best of class” learner in several domains~\cite{Rokach2016}, including fraud detection~\cite{BHATTACHARYYA2011}.

\smallskip
For the above classifiers, we leverage the implementation provided by the Weka machine learning library~\cite{WEKA2016}.

  \subsection{Performance measures and validation}
\label{subsec:performanceMeasures}
\label{subsec:validation}

To evaluate the performance of our detection models, 
we rely on best practices from the literature.

Specifically, in the context of binary problems with imbalanced class distributions, 
as in the case here considered, the rare class is denoted as the \textit{positive class}, 
while the majority class is denoted as the \textit{negative class}. 
The following terminology is then used to describe how the model performs on a given set of test instances: 
a \textit{true positive} (resp.\ \textit{negative}) is a positive (resp.\ negative) instance correctly classified by the model; 
a \textit{false negative} is a positive instance wrongly classified as negative; 
a \textit{false positive} is a negative instance wrongly classified as positive.

Depending on the specific characteristics of the data at hand, 
different metrics can be used for quantifying the extent to which 
the model is able to recognize positive and negative instances~\cite{WEKA2016}. 
Hereafter, TP (resp.\ TN) refers to the number of true positives (resp.\ negatives), 
while FP (resp.\ FN) refers to the number of false positives (resp.\ negatives):

\begin{LaTeXdescription}
	\item[Accuracy] ((TP + TN)/(TP + TN + FP + FN)) is the fraction of test instances whose class is predicted correctly;
	\item[Specificity] (TN/(TN+FP)) is the fraction of negative instances classified correctly;
	\item[Sensitivity] (TP/(TP+FN)), also called \textbf{Recall}, is the fraction of positive instances classified correctly;
	\item[Precision] (TP/(TP+FP)) is the fraction of instances that actually are positive in the group the model has predicted as positive;
	\item[F-measure] (2$\cdot$Precision$\cdot$Recall / (Precision+Recall)) is the harmonic mean between precision and recall;
	\item[G-mean] (Recall$\cdot$Specificity)$^{(0.5)}$ is the geometric mean between specificity and recall;
	\item[AUC] is the area under the \textit{Receiver Operating Characteristics} (ROC) curve, which shows the trade-off between true positive and false positive rates (the better the model, the closer the area is to 1).
\end{LaTeXdescription}

Accuracy is the most common performance metric but, alone, 
is not suited for evaluating models induced from imbalanced datasets. 
In a fraud detection context, if 0.5\% of the instances are fraudulent (as in our dataset),
then a model that predicts every instance as non-fraudulent has an accuracy of 99.5\%, 
even though it fails to detect any of the frauds. 
In this situation, class-specific metrics (such as specificity, recall and precision) 
can help to better describe and understand the model behaviour. 
In particular, recall and precision are widely used in applications 
where the successful detection of the rare class is considered more interesting or important 
(as for the ‘Ponzi’ class in our problem).

To avoid building models that maximize one metric at the expense of another, 
trade-off values such those expressed by F-measure and G-mean are taken into account. 
In turn, AUC is usually considered more significant than accuracy 
when comparing the overall performance of different classifiers.

Instead of using a single test set to compute the above metrics, 
we adopt an iterative \textit{cross-validation} protocol, 
which involves splitting the original dataset into $K$ subsets of the same size (\textit{folds}). 
At each iteration, one of the folds is retained as test data for evaluating the model performance, 
while the remaining $K-1$ folds are used as training data for building the model. 
This procedure is repeated $K$ times, using each time a different fold as test set, 
and the results of the $K$ runs are finally aggregated to obtain TP, TN, FP and FN counts. 
In our experiments, based on common practise in the literature, we set $K = 10$.

  \subsection{Results}
\label{sec:datamining-results}

We now present the results obtained with RIPPER, Bayes Net and Random Forest classifiers
presented in~\Cref{sec:datamining:classifiers}.
Hereafter, we use the following acronyms: \textit{RIP} for RIPPER, \textit{BN} for Bayes Net and \textit{RF} for Random Forest. 

First, we evaluate their performance without applying the sampling-based and the cost-sensitive approaches 
(see~\Cref{sec:datamining:imbalance}). 
The results are shown in~\Cref{fig:matricesResultsOriginaDataset} in the form of confusion matrices, 
where the row index refers to the actual class, while the column index refers to the predicted class. 
As we can see, the classification performance is not satisfactory, due to the high number of Ponzi schemes not recognized. 
Bayes Net classifies correctly the largest number of Ponzi instances (23 out of \totPonzi), 
but the number of false positives (\ie, non-Ponzi classified as Ponzi) is the highest as well. 
These results confirm that learning from highly imbalanced datasets is a very difficult task.

As a next step, we explore the effectiveness of the random undersampling approach, 
which has proved to be useful to deal with datasets where the fraud rate is comparable with the one here considered~\cite{BHATTACHARYYA2011}. 
Specifically, at each iteration of the cross-validation procedure, 
we manipulate the training set to reduce the extent of class imbalance: 
the original proportion of 1 Ponzi instance every 200 instances of non-Ponzi (1:200) 
is reduced to 1 Ponzi every 40 non-Ponzi (1:40), 1 Ponzi every 20 non-Ponzi (1:20), 
1 Ponzi every 10 non-Ponzi (1:10) and 1 Ponzi every 5 non-Ponzi (1:5). 
Note that we do not modify the class distribution of the test instances, 
to not introduce any bias in the final performance estimate.

The results in~\Cref{fig:matricesResultsUndersampling} 
show that the undersampling approach results in an improved true positive rate.
In particular, within the 1:5 setting, Random Forest recognizes 
the same number of Ponzi as Bayes Net (25 out of \totPonzi), 
but with a significantly lower number of false positives. 
Bayes Net, indeed, produces too many false positives (266), 
even more than RIPPER (226). 
Thus, while improving the true positive rate (and hence the recall metric), 
the undersampling approach is not quite satisfactory 
in terms of false positives (that affect the precision metric). 
This difficulty of achieving an optimal trade-off between recall and precision 
is a recognized issue in the fraud detection literature~\cite{ABDALLAH2016}.

As a further step, we investigate the effectiveness 
of the cost-sensitive approach. 
When learning our detection models, 
we use the cost matrices shown in~\Cref{fig:CostMatrices}. 
In the matrix \textit{CM5}, 
the cost of committing a false negative error is 5 times larger 
than the cost of committing a false positive error; 
it is 10 times larger in \textit{CM10}, 20 times larger in \textit{CM20} and 40 times larger in \textit{CM40}. 

The results achieved by the cost-sensitive classifiers 
are shown in~\Cref{fig:matricesResultCostMatrix}. 
As we can see from the confusion matrices, RIPPER obtains the same results
as in the original setting (\Cref{fig:matricesResultsOriginaDataset}). 
This is not surprising, since the algorithm is designed to cope with 
the rare (\ie, positive) class 
and turns out to be insensitive to further penalising a wrong classification of the positive instances. 
In turn, Bayes Net does not seem to take significant advantage 
of the cost-sensitive approach, 
which results in 24 true positives (one more than in the original setting), 
but with an increased number of false positives.

The best results are obtained with the Random Forest classifier. 
Using the \textit{CM5} matrix, it recognizes 25 Ponzi schemes, 
as Bayes Net in the 1:5 undersampling setting, 
but with a strong reduction of the false positives (only 13). 
When penalising more the false negatives, 
the number of the true positives increases 
(29 using \textit{CM10} and 31 using \textit{CM20}) 
and the number of false positives increases in turn, 
but to an acceptable extent (26 e 77 respectively). 
Increasing further the cost (\textit{CM40}) is not beneficial 
since the number of true positives remains the same but the false positives 
increase to 132. 

In~\Cref{fig:metricsResult}, 
we further detail the performance of the cost-sensitive Random Forest classifier,
that has shown to be superior to the other approaches here explored. 
Different performance metrics are computed as explained in~\Cref{subsec:performanceMeasures}).

In terms of accuracy, 
which simply expresses the fraction of correctly classified instances 
(irrespective of their class), 
the best result is achieved with the \textit{CM5} cost matrix. 
It also ensures the highest true negative rate (specificity) 
and a good trade-off between recall and precision (in terms of F-measure). 
However, given the specific characteristics of the considered domain, 
where the correct classification of Ponzi schemes is of paramount importance, 
we consider especially relevant the recall value, 
which is optimised using the \textit{CM20} cost matrix. 
The G-mean value, that expresses a trade-off between recall and specificity, 
is also better with \textit{CM20} and, in this setting, 
the AUC value is the highest as well.

Taking these considerations into account, 
the Random Forest model obtained using \textit{CM20} 
can be considered the most effective for detecting Ponzi schemes.


\begin{figure*}
	\centering
	\begin{tabular}{l|c|c|c|}
		\multicolumn{2}{c}{}&\multicolumn{2}{c}{Predicted}\\
		\cline{3-4}
		\multicolumn{1}{c}{}& RIP &\Ponzi&\nonPonzi\\
		\cline{2-4}
		\multirow{3}{*}{\rotatebox{90}{Actual}} 
		& \Ponzi & 19 & 13 \\
		\cline{2-4}& \nonPonzi & 7 & 6393\\
		\cline{2-4}
	\end{tabular}
	\begin{tabular}{l|c|c|c|}
	\multicolumn{2}{c}{}&\multicolumn{2}{c}{Predicted}\\
	\cline{3-4}
		\multicolumn{1}{c}{}& BN &\Ponzi&\nonPonzi\\
	\cline{2-4}
	\multirow{3}{*}{\rotatebox{90}{Actual}} 
	& \Ponzi & 23 & 9 \\
	\cline{2-4}& \nonPonzi & 99 & 6301\\
	\cline{2-4}
	\end{tabular}
	\begin{tabular}{l|c|c|c|}
	\multicolumn{2}{c}{}&\multicolumn{2}{c}{Predicted}\\
		\cline{3-4}
		\multicolumn{1}{c}{}& RF &\Ponzi&\nonPonzi\\
		\cline{2-4}
		\multirow{3}{*}{\rotatebox{90}{Actual}} 
		& \Ponzi & 11 & 21 \\
		\cline{2-4}& \nonPonzi & 0 & 6400\\
		\cline{2-4}
	\end{tabular} 
	\caption{Confusion matrices for RIPPER, Bayes Net and Random Forest.}
	\label{fig:matricesResultsOriginaDataset}
\end{figure*}

\begin{figure*}
  \centering
  
  \begin{tabular}{l|l|c|c|}
  	\multicolumn{2}{c}{}&\multicolumn{2}{c}{RIP: 1:40}\\
  	\cline{3-4}
  	\multicolumn{2}{c|}{}& \Ponzi & \nonPonzi \\
  	\cline{2-4}
  	& \Ponzi & 21 & 11 \\
  	\cline{2-4}& \nonPonzi & 44 & 6356\\
  	\cline{2-4}
  \end{tabular}
  \begin{tabular}{l|l|c|c|}
  	\multicolumn{2}{c}{}&\multicolumn{2}{c}{RIP: 1:20}\\
  	\cline{3-4}
  	\multicolumn{2}{c|}{}& \Ponzi & \nonPonzi \\
  	\cline{2-4}
  	& \Ponzi & 23 & 9 \\
  	\cline{2-4}& \nonPonzi & 66 & 6334\\
  	\cline{2-4}
  \end{tabular}
  \begin{tabular}{l|l|c|c|}
  	\multicolumn{2}{c}{}&\multicolumn{2}{c}{RIP: 1:10}\\
  	\cline{3-4}
  	\multicolumn{2}{c|}{}& \Ponzi & \nonPonzi \\
  	\cline{2-4}
  	& \Ponzi & 23 & 9 \\
  	\cline{2-4}& \nonPonzi & 97 & 6303\\
  	\cline{2-4}
  \end{tabular}
  \begin{tabular}{l|l|c|c|}
  	\multicolumn{2}{c}{}&\multicolumn{2}{c}{RIP: 1:5}\\
  	\cline{3-4}
  	\multicolumn{2}{c|}{}& \Ponzi & \nonPonzi \\
  	\cline{2-4}
  	& \Ponzi & 24 & 8 \\
  	\cline{2-4}& \nonPonzi & 226 & 6174\\
  	\cline{2-4}
  \end{tabular}
  \\
  \medskip
  \begin{tabular}{l|l|c|c|}
    \multicolumn{2}{c}{}&\multicolumn{2}{c}{BN: 1:40}\\
    \cline{3-4}
    \multicolumn{2}{c|}{}& \Ponzi & \nonPonzi \\
    \cline{2-4}
                        & \Ponzi & 23 & 9 \\
    \cline{2-4}& \nonPonzi & 154 & 6246\\
    \cline{2-4}
  \end{tabular}
  \begin{tabular}{l|l|c|c|}
    \multicolumn{2}{c}{}&\multicolumn{2}{c}{BN: 1:20}\\
    \cline{3-4}
    \multicolumn{2}{c|}{}& \Ponzi & \nonPonzi \\
    \cline{2-4}
                        & \Ponzi & 23 & 9 \\
    \cline{2-4}& \nonPonzi & 185 & 6215\\
    \cline{2-4}
  \end{tabular}
  \begin{tabular}{l|l|c|c|}
    \multicolumn{2}{c}{}&\multicolumn{2}{c}{BN: 1:10}\\
    \cline{3-4}
    \multicolumn{2}{c|}{}&\Ponzi&\nonPonzi\\
    \cline{2-4}
                        & \Ponzi & 24 & 8 \\
    \cline{2-4}& \nonPonzi & 233 & 6167\\
    \cline{2-4}
  \end{tabular}
  \begin{tabular}{l|l|c|c|}
    \multicolumn{2}{c}{}&\multicolumn{2}{c}{BN: 1:5}\\
    \cline{3-4}
    \multicolumn{2}{c|}{}&\Ponzi&\nonPonzi\\
    \cline{2-4}
                        & \Ponzi & 25 & 7 \\
    \cline{2-4}& \nonPonzi & 266 & 6134\\
    \cline{2-4}
  \end{tabular}
  \\	
  \medskip
  \begin{tabular}{l|l|c|c|}
  	\multicolumn{2}{c}{}&\multicolumn{2}{c}{RF: 1:40}\\
  	\cline{3-4}
  	\multicolumn{2}{c|}{}&\Ponzi&\nonPonzi\\
  	\cline{2-4}
  	& \Ponzi & 17 & 15 \\
  	\cline{2-4}& \nonPonzi & 3 & 6397\\
  	\cline{2-4}
  \end{tabular}
  \begin{tabular}{l|l|c|c|}
  	\multicolumn{2}{c}{}&\multicolumn{2}{c}{RF: 1:20}\\
  	\cline{3-4}
  	\multicolumn{2}{c|}{}&\Ponzi&\nonPonzi\\
  	\cline{2-4}
  	& \Ponzi & 19 & 13 \\
  	\cline{2-4}& \nonPonzi & 9 & 6391\\
  	\cline{2-4}
  \end{tabular}
  \begin{tabular}{l|l|c|c|}
  	\multicolumn{2}{c}{}&\multicolumn{2}{c}{RF: 1:10}\\
  	\cline{3-4}
  	\multicolumn{2}{c|}{}&\Ponzi&\nonPonzi\\
  	\cline{2-4}
  	& \Ponzi & 21 & 11 \\
  	\cline{2-4}& \nonPonzi & 30 & 6370\\
  	\cline{2-4}
  \end{tabular}
  \begin{tabular}{l|l|c|c|}
  	\multicolumn{2}{c}{}&\multicolumn{2}{c}{RF: 1:5}\\
  	\cline{3-4}
  	\multicolumn{2}{c|}{}&\Ponzi&\nonPonzi\\
  	\cline{2-4}
  	& \Ponzi & 25 & 7 \\
  	\cline{2-4}& \nonPonzi & 70 & 6330\\
  	\cline{2-4}
  \end{tabular}
   \caption{Confusion matrices for RIPPER, Bayes Net and Random Forest across different undersampling in training data.}
  \label{fig:matricesResultsUndersampling}
\end{figure*}

\begin{figure*}
	\centering 
	\begin{tabular}{l|c|c|c|}
		\multicolumn{2}{c}{}&\multicolumn{2}{c}{Predicted}\\
		\cline{3-4}		
		\multicolumn{1}{c}{}& CM5 &\Ponzi&\nonPonzi\\
		\cline{2-4}
		\multirow{3}{*}{\rotatebox{90}{Actual}} 
		& \Ponzi & 0 & 5 \\
		\cline{2-4}& \nonPonzi & 1 & 0\\
		\cline{2-4}
	\end{tabular}
	\begin{tabular}{l|c|c|c|}
		\multicolumn{2}{c}{}&\multicolumn{2}{c}{Predicted}\\
		\cline{3-4}
		\multicolumn{1}{c}{}& CM10 &\Ponzi&\nonPonzi\\
		\cline{2-4}
		\multirow{1}{*}{\rotatebox{90}{Actual}} 
		& \Ponzi & 0 & 10 \\
		\cline{2-4}& \nonPonzi & 1 & 0\\
		\cline{2-4}
	\end{tabular}
	\begin{tabular}{l|c|c|c|}
		\multicolumn{2}{c}{}&\multicolumn{2}{c}{Predicted}\\
		\cline{3-4}
		\multicolumn{1}{c}{}& CM20 &\Ponzi&\nonPonzi\\
		\cline{2-4}
		\multirow{3}{*}{\rotatebox{90}{Actual}} 
		& \Ponzi & 0 & 20 \\
		\cline{2-4}& \nonPonzi & 1 & 0\\
		\cline{2-4}
	\end{tabular}
	\begin{tabular}{l|c|c|c|}
		\multicolumn{2}{c}{}&\multicolumn{2}{c}{Predicted}\\
		\cline{3-4}
		\multicolumn{1}{c}{}& CM40 &\Ponzi&\nonPonzi\\
		\cline{2-4}
		\multirow{1}{*}{\rotatebox{90}{Actual}} 
		& \Ponzi & 0 & 40 \\
		\cline{2-4}& \nonPonzi & 1 & 0\\
		\cline{2-4}
	\end{tabular}
	\caption{Cost matrices: CM5, CM10, CM20, and CM40.}
	\label{fig:CostMatrices}
\end{figure*}

\begin{figure*}
  \centering  	 
	\begin{tabular}{l|l|c|c|}
		\multicolumn{2}{c}{}&\multicolumn{2}{c}{RIP: CM5}\\
		\cline{3-4}
		\multicolumn{2}{c|}{}&\Ponzi&\nonPonzi\\
		\cline{2-4}
		& \Ponzi & 19 & 13 \\
		\cline{2-4}& \nonPonzi & 7 & 6393\\
		\cline{2-4}
	\end{tabular}
	\begin{tabular}{l|l|c|c|}
		\multicolumn{2}{c}{}&\multicolumn{2}{c}{RIP: CM10}\\
		\cline{3-4}
		\multicolumn{2}{c|}{}&\Ponzi&\nonPonzi\\
		\cline{2-4}
		& \Ponzi & 19 & 13 \\
		\cline{2-4}& \nonPonzi & 7 & 6393\\
		\cline{2-4}
	\end{tabular}
	\begin{tabular}{l|l|c|c|}
		\multicolumn{2}{c}{}&\multicolumn{2}{c}{RIP: CM20}\\
		\cline{3-4}
		\multicolumn{2}{c|}{}&\Ponzi&\nonPonzi\\
		\cline{2-4}
		& \Ponzi & 19 & 13 \\
		\cline{2-4}& \nonPonzi & 7 & 6393\\
		\cline{2-4}
	\end{tabular}
	\begin{tabular}{l|l|c|c|}
		\multicolumn{2}{c}{}&\multicolumn{2}{c}{RIP: CM40}\\
		\cline{3-4}
		\multicolumn{2}{c|}{}&\Ponzi&\nonPonzi\\
		\cline{2-4}
		& \Ponzi & 19 & 13 \\
		\cline{2-4}& \nonPonzi & 7 & 6393\\
		\cline{2-4}
	\end{tabular}
  \\
  \medskip 
  \begin{tabular}{l|l|c|c|}
    \multicolumn{2}{c}{}&\multicolumn{2}{c}{BN: CM5}\\
    \cline{3-4}
    \multicolumn{2}{c|}{}&\Ponzi&\nonPonzi\\
    \cline{2-4}
                        & \Ponzi & 24 & 8 \\
    \cline{2-4}& \nonPonzi & 136 & 6264\\
    \cline{2-4}
  \end{tabular}
  \begin{tabular}{l|l|c|c|}
    \multicolumn{2}{c}{}&\multicolumn{2}{c}{BN: CM10}\\
    \cline{3-4}
    \multicolumn{2}{c|}{}&\Ponzi&\nonPonzi\\
    \cline{2-4}
                        & \Ponzi & 24 & 8 \\
    \cline{2-4}& \nonPonzi & 155 & 6245\\
    \cline{2-4}
  \end{tabular}
  \begin{tabular}{l|l|c|c|}
    \multicolumn{2}{c}{}&\multicolumn{2}{c}{BN: CM20}\\
    \cline{3-4}
    \multicolumn{2}{c|}{}&\Ponzi&\nonPonzi\\
    \cline{2-4}
                        & \Ponzi & 24 & 8 \\
    \cline{2-4}& \nonPonzi & 192 & 6203\\
    \cline{2-4}
  \end{tabular}
  \begin{tabular}{l|l|c|c|}
    \multicolumn{2}{c}{}&\multicolumn{2}{c}{BN: CM40}\\
    \cline{3-4}
    \multicolumn{2}{c|}{}&\Ponzi&\nonPonzi\\
    \cline{2-4}
                        & \Ponzi & 24 & 8 \\
    \cline{2-4}& \nonPonzi & 213 & 6187\\
    \cline{2-4}
  \end{tabular}
  \\
  \medskip
  \begin{tabular}{l|l|c|c|}
  	\multicolumn{2}{c}{}&\multicolumn{2}{c}{RF: CM5}\\
  	\cline{3-4}
  	\multicolumn{2}{c|}{}&\Ponzi&\nonPonzi\\
  	\cline{2-4}
  	& \Ponzi & 25 & 7 \\
  	\cline{2-4}& \nonPonzi & 13 & 6387\\
  	\cline{2-4}
  \end{tabular}
  \begin{tabular}{l|l|c|c|}
  	\multicolumn{2}{c}{}&\multicolumn{2}{c}{RF: CM10}\\
  	\cline{3-4}
  	\multicolumn{2}{c|}{}&\Ponzi&\nonPonzi\\
  	\cline{2-4}
  	& \Ponzi & 29 & 3 \\
  	\cline{2-4}& \nonPonzi & 26 & 6374\\
  	\cline{2-4}
  \end{tabular}
  \begin{tabular}{l|l|c|c|}
  	\multicolumn{2}{c}{}&\multicolumn{2}{c}{RF: CM20}\\
  	\cline{3-4}
  	\multicolumn{2}{c|}{}&\Ponzi&\nonPonzi\\
  	\cline{2-4}
  	& \Ponzi & 31 & 1 \\
  	\cline{2-4}& \nonPonzi & 77 & 6323\\
  	\cline{2-4}
  \end{tabular}
  \begin{tabular}{l|l|c|c|}
  	\multicolumn{2}{c}{}&\multicolumn{2}{c}{RF: CM40}\\
  	\cline{3-4}
  	\multicolumn{2}{c|}{}&\Ponzi&\nonPonzi\\
  	\cline{2-4}
  	& \Ponzi & 31 & 1 \\
  	\cline{2-4}& \nonPonzi & 132 & 6268\\
  	\cline{2-4}
  \end{tabular}
  \caption{Confusion matrices of RIPPER, Bayes Net and Random Forest across different cost-matrices.}
  \label{fig:matricesResultCostMatrix}
\end{figure*}

\begin{figure*}
  \centering
  \begin{tabular}{l c c c c c c c}
    \cline{1-8}
    Random Forest & Accuracy & Recall & Specificity & F-measure & Precision & G-mean & AUC \\
    \cline{1-8}
    CM5 : Using Cost 5  & .997 & .781 & .998 & .714 & .658 & .883 & .890\\
    CM10: Using Cost 10 & .995 & .906 & .995 & .667 & .527 & .949 & .951\\
    CM20: Using Cost 20 & .988 & .969 & .987 & .443 & .287 & .978 & .978\\
    CM40: Using Cost 40 & .979 & .969 & .979 & .318 & .190 & .973 & .974\\
    \cline{1-8}
  \end{tabular}
  \caption{Performance of Random Forest across different cost-matrices.}
  \label{fig:metricsResult}
\end{figure*}

  \subsection{Application of the induced model}
\label{sec:datamining:application}

In this~\namecref{sec:datamining:application} perform an \emph{ex-post} 
validation of the best classifier obtained so far, \ie Random Forest with \textit{CM20}.
To this purpose we collect other Ponzi schemes by searching the web,
with the same methodology of~\Cref{sec:dataset:collection}.
We report their addresses in~\Cref{fig:dataset:collectionNewPonzi}. 
Overall, the 20 Ponzi schemes in this collection
have gathered more than $15$ millions \USD, 
in large part with a single scheme, CryptoSplit 
(see~\Cref{fig:newdataset:clustering}).

We then construct an alternative dataset, 
comprising the features of the clusters of the Ponzi schemes
in the new collection,
and those of 4000 randomly-chosen Bitcoin addresses 
not in the original dataset.
By applying the Random Forest classifier with \textit{CM20}
to the alternative dataset, 
we obtain the following confusion matrix:

\begin{center}
  \begin{tabular}{l|l|c|c|}
    \cline{3-4}
    \multicolumn{2}{c|}{}&\Ponzi&\nonPonzi\\
    \cline{2-4}
                        & \Ponzi & 18 & 2 \\
    \cline{2-4}& \nonPonzi & 81 & 3919\\
    \cline{2-4}
  \end{tabular}
\end{center}

\smallskip
Notably, the classifier recognizes 18 Ponzi schemes out of 20, 
producing 81 false positives.
The 2 Ponzi schemes not recognized by the classifier are marked
with $\xmark$ in~\Cref{fig:newdataset:clustering}.

\begin{table}[t!]
  \centering				
  \small
  \caption{Addresses of the alternative set of Ponzi schemes.}
  \resizebox{\columnwidth}{!}{
    \begin{tabular}{ll}  
      \toprule
      \textbf{Ponzi scheme} & \textbf{Deposit address} \\
      \midrule		
      Longtermpaying &	1MnuUkqvsyZdwd3xyM354kqVoPBhfBGE78  \\
      ebitinvest.com &	1LkT3qubANxtSHvxokZ8Nkrv6k7EFi6F1 \\
      PonziCoin &	1NcHirWVDfUAngWLjBzmPCQaeZaMPCceHC \\
      CoinDoubleP2 &	15vr3X25cgfMBXpX8PQP3M6bQViFgqrm6U \\
      CoinDoubleP1 &	1AA4A7cbVf3wMtG1RrhDoPkjAX2C1RJMjW \\
      CoinDoubleP3 &	1NNSgNDU52W79QbXHGSBYHW874nQYb7oms \\
      CryptoSplit2 &	1P5rm8YmufwfNdqg6Dy47boaeBCXvEDjUP \\
      CryptoSplit	 &147MddkTvgHR2kEoEpj5fjx7MK71va54y5 \\
      TrustedBusinessInvestments  &	1NWdUDU4X91JTKEFJRKgmW4yYsUhWnaMJH \\
      SmallProfit	 &133ySbYkiA7BTtau2v3Hs4GLoDgGZFNDbD \\
      SmallProfit (2)	 &1Bb4JG51DizK6iSn4w4RqhRNXPUpkhacHx \\
      MagicBitcoinDoubler  &	1QLbGuc3WGKKKpLs4pBp9H6jiQ2MgPkXRp \\
      coin-generator.net &	18Xiqg52FfgA43rqCyCU5iqq6KNBgjTBj8 \\
      MMM Global &	1MxA5W1TKcMwLNh6EYL9QwAMLXtifHnxwb \\
      BTC-flow (2) &	1CjGx5ujxvzdbZqzzhPREXqvxoYSgDoAgd \\
      WeeklyPonzi &	1CHJArco4Qv6cmTZNF7Km7cuATCD1Z1NSu \\
      DoubleBot &	1LWadswFVXwCoVSKAeo3tuxWKqKr1EWFxR \\
      ClearHash.net (1) &	1AFjgfnUhAYp4eh2GhbbLkCXY5xK25qJmQ \\
      ClearHash.net (2) &	1DXxLzocfWXTHVYv4MTai4LLtXZgcJDknZ \\
      ClearHash.net (4) &	19g9exzmtJ2sQbBBB3x2PiY9pReVCm8HqA \\
      \bottomrule
    \end{tabular}
  } 
  \label{fig:dataset:collectionNewPonzi}
\end{table}

\begin{table}[t!]
  \centering				
  \small
  \caption{Alternative set of Ponzi schemes, by cluster size.}
  \resizebox{\columnwidth}{!}{
    \begin{tabular}{lrrrrrr}
      \toprule
      \textbf{Ponzi scheme} & \textbf{{\#}Addr.} & \textbf{{\#}Tx}
      & \textbf{In (\BTC)} & \textbf{In (\$)}  \\ 
      \midrule
      CryptoSplit          & 1763 & \USDfmt{126245} & 35,654 & \USDfmt{15124204}{}\\
      PonziCoin            & 243  & \USDfmt{3226}   & 229    & \USDfmt{132158}{}  \\
      MagicBitcoinDoubler  & 135  & \USDfmt{49239}  & 404    & \USDfmt{77049}{}   \\
      coin-generator.net   & 36   & \USDfmt{468}    & 2      & \USDfmt{3608}{}    \\
      TrustedBusinessInvest& 22   & \USDfmt{91}     & 4      & \USDfmt{11361}{}   \\ 
      DoubleBot            & 21   & \USDfmt{298}    & 0.35   & \USDfmt{97}{}      \\
      WeeklyPonzi          & 5    & \USDfmt{827}    & 13     & \USDfmt{6724}{}    \\
      CoinDoubleP1 $\xmark$ & 3    & \USDfmt{90}     & 3      & \USDfmt{1422}{}    \\
      CoinDoubleP3         & 2    & \USDfmt{10}     & 4      & \USDfmt{1680}{}    \\
      CoinDoubleP2         & 2    & \USDfmt{167}    & 3      & \USDfmt{1238}{}    \\
      CryptoSplit2         & 2    & \USDfmt{292}    & 1      & \USDfmt{336}{}     \\			ClearHash.net (2)    & 2    & \USDfmt{30}     & 0.5    & \USDfmt{142}{}     \\			BTC-flow (2)         & 2    & \USDfmt{308}    & 0.08   & \USDfmt{34}{}      \\			ClearHash.net (4)    & 2    & \USDfmt{13}     & 0.01   & \USDfmt{5}{}       \\			SmallProfit          & 1    & \USDfmt{8518}   & 32     & \USDfmt{19961}{}   \\
      ebitinvest.com       & 1    & \USDfmt{414}    & 16     & \USDfmt{39861}{}   \\
      Longtermpaying       & 1    & \USDfmt{1504}   & 13     & \USDfmt{3698}{}    \\
      MMM Global $\xmark$  & 1    & \USDfmt{22}     & 7      & \USDfmt{1663}{}    \\
      ClearHash.net (1)    & 1    & \USDfmt{801}    & 5      & \USDfmt{2094}{}    \\
      SmallProfit (2)      & 1    & \USDfmt{1991}   & 0.29   & \USDfmt{185}{}     \\
      
      \midrule 
      \textbf{Total (20 schemes)} & \USDfmt{2246} & \USDfmt{194515} & \BTCfmt{36398} & \USDfmt{15427373}{} \\
      \bottomrule 
    \end{tabular}
  }
  \label{fig:newdataset:clustering}
\end{table}

  \subsection{Ranking and evaluation of features}
\label{sec:datamining:ranking}

We now study which features, among those listed in~\Cref{sec:dataset:features},
are more relevant for the classification of Ponzi schemes.
To this purpose we exploit the feature selection functionality of 
Weka~\cite{WEKA2016}, which implements several methods for ranking features.
Among them, we apply some univariate methods 
(\emph{Information Gain}, \emph{Gain Ratio}, \emph{Symmetrical Uncertainty},
and \emph{OneR}),
and the multivariate method \emph{ReliefF}.

Among the \totFeatures features included in our datasets,
we consider those with the highest number of occurrences 
in the first positions of these rankings,
and thus can be considered as the most discriminating ones.
These features are the following:
\begin{inlinelist}

\item the Gini coefficient of the outgoing values;

\item the ratio between incoming and total transactions;
  
\item the average and standard deviation of the outgoing values;
  
\item the number of different addresses who have transferred money to 
the cluster, and subsequently received money from it;

\item the lifetime of the cluster, and the number of activity days.
 
\end{inlinelist}

\section{Conclusions}
\label{sec:conclusions}

Criminal activities that 
accept payments in bitcoins damage the reputation of Bitcoin, 
and eventually may be detrimental to the diffusion of cryptocurrencies
for legitimate uses.
However, since all currency transfers are recorded on a public ledger, 
surveillance authorities can analyse them, 
trying to detect anomalous or suspect behaviours.

Despite the transparency of the blockchain,
tracking illicit financial flows is a challenging problem,
for several reasons.
First, many illicit activities involve hundreds, or even thousands,
of transactions --- thus making manual inspection impracticable.
For instance, the Ponzi schemes in our dataset use ${\sim}1400$ 
transactions on average: this number is sufficiently large 
to discourage any attempt at manual inspection.
Second, the number of illicit flows is overwhelmed by that of
legitimate ones, making the task of surveillance authorities
similar to finding ``the needle in the haystack''.
Another difficulty is that smart cyber-criminals exploit techniques 
to make analysing their activities more difficult,
\eg by using mixing services to hide the actual provenance of illegal money.
All these observations highlight the pressing need 
for automated techniques to detect illegal activities on cryptocurrencies.

In this work we have proposed an automatic analysis of 
Ponzi schemes on Bitcoin, based on supervised learning algorithms.
Ponzi schemes are a classic fraud masqueraded as ``high-yield'' 
investment schemes.
However, in a Ponzi scheme the investors are repaid only with the 
funds invested by new users, hence eventually the scheme implodes, 
as at a certain point it will no longer be possible to find new investments.

After a preliminary phase of manual search, 
we have identified \totPonzi Bitcoin addresses used by Ponzi schemes.
Address clustering allowed us to
extend our collection to \totCluster addresses, 
which overall received investments for \mbox{$\sim 10$} millions \USD.
We have devised a set of features of clusters, 
that we have used to create a dataset containing 
the features of all the addresses of Ponzi schemes,
and those of other \totRandomAddresses randomly-chosen addresses.
We experimented with data mining tools to evaluate different
supervised learning strategies.
The best classifier we have found correctly classifies 31 Ponzi schemes
out of \totPonzi, producing $\sim 1\%$ of false positives. 

An obvious extension of this work is to apply this classifier
to \emph{all} the addresses in the Bitcoin blockchain. 
This kind of analysis poses serious efficiency issues, 
since several dozens of millions of distinct Bitcoin addresses 
have been used so far.
Although the number of false positives is quite low 
(comparable to that of other successful approaches in the 
fraud detection literature~\cite{BHATTACHARYYA2011}),
automated techniques to check the false positives are in order.
To this purpose one could exploit auxiliary information sources,
\eg web discussion forums, 
and the IP addresses collected by monitoring
the traffic on the Bitcoin network.

Our classifier can also be used to detect Ponzi schemes 
implemented over other cryptocurrencies, like \eg Ethereum. 
To this purpose we could exploit public datasets 
of Ethereum-based Ponzi schemes~\cite{Bartoletti17dissecting},
which collect addresses and other relevant data of 152 Ponzi schemes.
In the case of Ethereum, the precision of the classifier could
be improved by exploiting more specific features, 
like \eg the distribution of gain among users,
and the correlation between the timings of inflows and outflows
observed in~\cite{Bartoletti17dissecting}.

The approach we have followed in this work can be exploited 
for the detection of other cryptocurrency-based frauds besides Ponzi schemes,  
like \eg ransomware, money laundering, \etc
This would require, as in our case, 
a preliminary phase of dataset construction.
The dataset could take into account, 
besides the features used in~\Cref{sec:dataset:features}, 
further features that better capture specific behaviours 
of the fraud under analysis.

A relevant question is what interventions can be devised 
after an illegal activity has been detected.
The \emph{ex-post} sanitization of fraudulent activities is hampered
by the current \emph{fungibility} of the Bitcoin currency.
This means that Bitcoin users and exchanges are not selective
in which bitcoins to accept, and which ones to reject.
Hence, even if we set up risk scores for Bitcoin transactions
as proposed in~\cite{Moser14fc},
\eg by marking as ``bad'' all the bitcoins flowing out a Ponzi scheme,
it would not be possible to take countermeasures to the use of ``bad'' bitcoins
until they leave the Bitcoin ecosystem through an exchange service.

\bibliographystyle{IEEEtran}
\bibliography{main} 

\newpage
\setcounter{tocdepth}{1}
\listoffixmes

\end{document}